# Accelerating grain boundary modelling and simulation by automated pre-evaluation of the irreducible macroscopic representation

## Author information

Wei Wan

*Department of Materials Science and Engineering, City University of Hong Kong, Hong Kong*

*Email address:* vanv@email.ncu.edu.cn

## Abstract

A major challenge in the modelling and simulation of grain boundaries (GBs) is the conflict between the system size and computation capability, which inherently restricts the computationally accessible boundary characters. We present an algorithm to find the irreducible macroscopic representation of any given coincident-site-lattice GB character in the cubic lattice, and thus determine the irreducible size of its supercell. The algorithm is compared with the conventional orthogonal supercell and a published calculation method to assess its merits in saving computational resources. This supercell size parameter can be used to predict which GB character is relatively special across the vast 5D space, as those GBs possess small structural units are likely to exhibit particular structure-property relationships that are worthy of attention. The prediction is confirmed by examining the energy and mobility trends of aluminum mixed GBs obtained from the atomistic simulations.

## Introduction

Grain boundaries (GBs) are interfaces separating individual crystals and affecting many properties in polycrystalline materials. For example, the strength [1–3], ductility [4], diffusion [5], corrosion resistance [6], thermal and electronic conductivities [7, 8] can be improved by controlling the population of the desired GB types, which is also known as GB engineering [9]. To enhance the material properties via GB engineering [10–13], we need to know how GBs change the performance at the basic structural aspect. In other words, the structure-property relationship over a range of different GBs should be comprehended [14–16].

The structure of a GB is primarily defined by its five macroscopic degrees of freedom (DOFs). The macroscopic DOFs describe the rotation between two identical grains and are often referred to as the GB character. For each unique GB character, there is a multiplicity of microscopic DOFs where the GB contains atoms with different numbers and positions. Such multiplicity is typically represented by metastable states or GB phases [17–20], since varying the atomic arrangements could result in structures with different energies. The minimum energy structure always receives the most attention due to its thermodynamical stability, and those metastable states are also important as they provide information on the finite temperature equilibrium and non-equilibrium properties [21].

The GB character also determines whether its structure is simple or complex. For instance, the GB characters can be divided into tilt, twist, and mixed (tilt-twist) types based on the geometry relationship between the boundary plane and the rotation axis, or low-angle and high-angle types based on the rotation (disorientation) angle [22]. Low angle tilt GBs are often an array of edge dislocations, in contrast to the low angle mixed GBs comprised of a complex dislocation network [23]. On the other hand, high angle tilt and twist GBs usually possess lower excess energy than the mixed types [24]. For the simple GBs, experimental tools such as high-resolution transmission electron microscopy (HRTEM) successfully reported their structures and migrations [25–28], but the inherent structural complexity of the mixed GB character hinders the expansion of such success [25].

Therefore, modelling and simulation become an alternative option in the investigation of the various GB structure-property relationships. The primary simulation methods include first-principles calculation and atomistic simulation. The former is relatively accurate and often used to resolve the electronic properties for small GB systems at around $10^2$ atoms [29]; but the latter could model large GB systems with atoms up to $10^6$, based on various empirical interatomic potentials [30]. This system size allows the atomistic simulations to capture not only the energetics but also the kinetic properties such as mobility and shear-coupling [31]. It should be noted that even the high-throughput atomistic simulations must capture the vast 5D character space at a certain resolution, and represent it with finite coincident-site-lattice (CSL) GB characters [32]. From the simulations and available experimental results, many methods have been developed to describe different GB structures at the atomic level, for instance, the dislocation model [33], structural unit model [34, 35], polyhedral unit model [36], and some metrics based on local atomic environments [37-39]. More recently, with the improvement of simulation techniques, fruitful results [24, 32, 40–42] about different GB types have started to emerge, though the computational accessibility of those mixed GB types is still limited by their complexity.

To facilitate the emerging understanding of different GB structure-property relationships, their structures must be constructed first. The available software tools for this target are numerous, including *GBstudio* [43], *CrystalMaker* [44], *Aimsgb* [45], and even *Atomsk* [46]. However, to the author's knowledge, no such tool can provide a solution capable of increasing accessible GB structures and reducing the computational cost simultaneously.

In this work, we propose an algorithm to find the irreducible macroscopic representation of a given CSL GB character, which could effectively expand the range of accessible GBs and increase the computational efficiency compared with the conventional approaches. By combining this algorithm with another sampling strategy, we build an automated procedure to minimize human intervention in the high-throughput GB sampling. Most importantly, it is revealed that the algorithm could determine which GB is relatively more special than the others, and should be considered in the investigation of the GB structure-property relationships.

## Methodology & discussion

### Algorithm procedure

The algorithm presented here is a basic version and limited to the CSL GBs in the pure material of the cubic lattice (e.g., FCC metals). Let's assume a case where the user has already obtained a CSL GB character via the existing construction/analytical metrics [24, 47–50]. For a given CSL GB character, the algorithm requires the user to convert the character to the form of *a pair of integer transformation matrices* $\{\boldsymbol{P}_\mathrm{T}, \boldsymbol{Q}_\mathrm{T}\}$ as its input following:

$$\boldsymbol{P}_\mathrm{T} = \begin{bmatrix} \boldsymbol{x}^p \\ \boldsymbol{y}^p \\ \boldsymbol{z}^p \end{bmatrix} = \begin{bmatrix} H_x^p & K_x^p & L_x^p \\ H_y^p & K_y^p & L_y^p \\ H_z^p & K_z^p & L_z^p \end{bmatrix}, \tag{1.1}$$

$$\boldsymbol{Q}_\mathrm{T} = \begin{bmatrix} \boldsymbol{x}^q \\ \boldsymbol{y}^q \\ \boldsymbol{z}^q \end{bmatrix} = \begin{bmatrix} H_x^q & K_x^q & L_x^q \\ H_y^q & K_y^q & L_y^q \\ H_z^q & K_z^q & L_z^q \end{bmatrix}, \tag{1.2}$$

Where superscripts $p$ and $q$ represent the ID of the transformation matrices of two grains. Integer vectors $\boldsymbol{x}$, $\boldsymbol{y}$ and $\boldsymbol{z}$ are the lattice orientations along the $x$, $y$ and $z$ (three orthogonal and right-handed) directions of the grain, respectively. Integers $H$, $K$ and $L$ are the Miller indices. Assuming the boundary plane is the $xy$ plane and the normal is along $z$ direction, then a CSL character requires $|\boldsymbol{x}^p| = |\boldsymbol{x}^q|$ and $|\boldsymbol{y}^p| = |\boldsymbol{y}^q|$. The mostly used orthogonal representation defines the 2D orthogonal size $S_\mathrm{O}$ of the GB supercell as:

$$S_\mathrm{O} = L_x \times L_y = a|\boldsymbol{x}^p| \times a|\boldsymbol{y}^p| \tag{2}$$

to satisfy the periodic boundary conditions in the two orthogonal ($x$ and $y$) directions within the boundary plane. Where $a$ is the lattice constant of the GB materials. The $z$ supercell length $L_z$ is designated by the user. In this case, there is a problem that $S_\mathrm{O}$ could be so large to make the structural iteration (to search the minimum energy structure) costly. For example, low angle GBs and the majority of the high angle mixed GBs are likely to have large values of $S_\mathrm{O}$, and thus taking them into account in the structure-property relationship is sometimes unbearable [23].

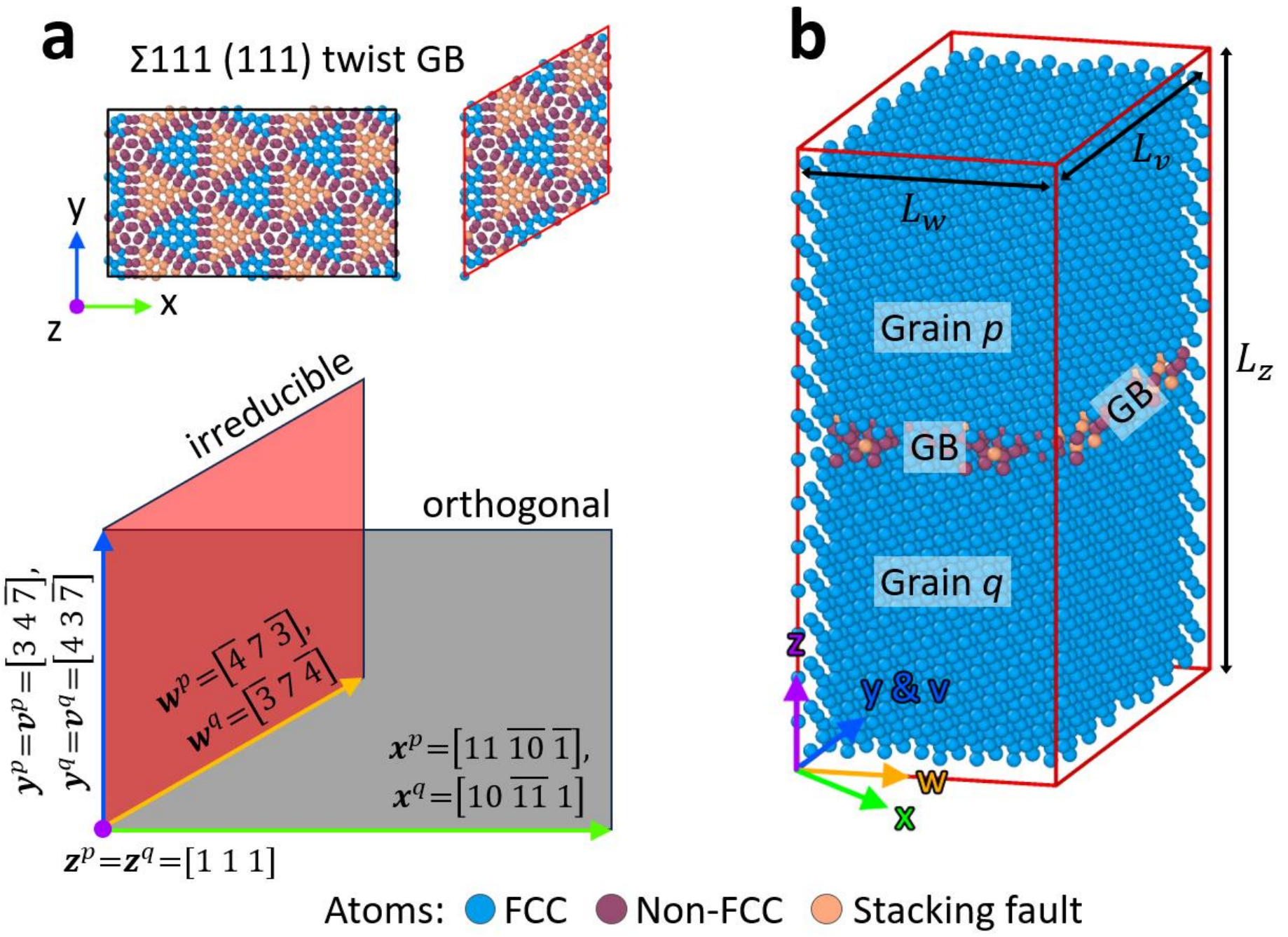

Figure 1. (a) Atomic structure of a Σ111 (111) twist GB (disorientation angle is 9.43°) under the orthogonal and irreducible macroscopic representations. (b) 3D snapshot of the GB supercell and its geometry definitions.

To solve the problem, the geometrical shape of the representation could be *monoclinic* (parallelogram) instead of orthogonal (rectangle). For example, in Figure 1a, using the monoclinic supercell has effectively reduced the system size for an exemplar Σ111 (111) twist GB, compared with the orthogonal supercell. For each unique GB character, it could be possible to find two pairs of lattice orientations (four linearly independent integer vectors) $\{\boldsymbol{w}^p, \boldsymbol{w}^q\}$ and $\{\boldsymbol{v}^p, \boldsymbol{v}^q\}$ respectively in the $\boldsymbol{x}^p\boldsymbol{y}^p$ and $\boldsymbol{x}^q\boldsymbol{y}^q$ planes, which define two 2D periodic supercells for grains $p$ and $q$, as well as a new irreducible GB supercell (c.f. Figure 1b) via coinciding the former two. Note that the four new lattice orientations should satisfy the following three constraints:

(I) The CSL constraint, which requires

$$|\boldsymbol{w}^p| = |\boldsymbol{w}^q|, |\boldsymbol{v}^p| = |\boldsymbol{v}^q|. \\ \boldsymbol{w}^p \nparallel \boldsymbol{w}^q \nparallel \boldsymbol{v}^p \nparallel \boldsymbol{v}^q. \tag{3.1}$$

(II) The rotation invariance constraint, which requires

$$\boldsymbol{w}^p\boldsymbol{z}^p = 0, \boldsymbol{w}^q\boldsymbol{z}^q = 0, \boldsymbol{v}^p\boldsymbol{z}^p = 0, \boldsymbol{v}^q\boldsymbol{z}^q = 0. \\ \mathrm{atan2}(\boldsymbol{w}^p \times \boldsymbol{x}^p, \boldsymbol{w}^p \cdot \boldsymbol{x}^p) = \mathrm{atan2}(\boldsymbol{w}^q \times \boldsymbol{x}^q, \boldsymbol{w}^q \cdot \boldsymbol{x}^q), \\ \mathrm{atan2}(\boldsymbol{v}^p \times \boldsymbol{x}^p, \boldsymbol{v}^p \cdot \boldsymbol{x}^p) = \mathrm{atan2}(\boldsymbol{v}^q \times \boldsymbol{x}^q, \boldsymbol{v}^q \cdot \boldsymbol{x}^q). \tag{3.2}$$

(III) The irreducible representation constraint, which requires

$$\max(|\mathrm{red}(\boldsymbol{w}^p)|, |\mathrm{red}(\boldsymbol{w}^q)|) < \max(|\mathrm{red}(\boldsymbol{v}^p)|, |\mathrm{red}(\boldsymbol{v}^q)|) < \max(|\mathrm{red}(\boldsymbol{u}^p)|, |\mathrm{red}(\boldsymbol{u}^q)|) < \cdots. \tag{3.3}$$

Where $\{\boldsymbol{u}^p, \boldsymbol{u}^q\}$ and … denote other candidate pairs of lattice orientations that satisfy the equations (3.1) and (3.2). red() is the operation to rescale the elements in an integer vector to their simplest ratio.

To find the required pairs of lattice orientations, we rotate $\boldsymbol{x}^p$ and $\boldsymbol{x}^q$ simultaneously about the vectors $\boldsymbol{z}^p$ and $\boldsymbol{z}^q$, respectively for $N$ times, to create a list of candidate pairs $\{\boldsymbol{u}_m^p, \boldsymbol{u}_m^q\}$:

$$\boldsymbol{u}_m^p = \mathrm{red}(\alpha_m\boldsymbol{x}^p + \beta_m\boldsymbol{y}^p), \boldsymbol{u}_m^q = \mathrm{red}(\alpha_m\boldsymbol{x}^q + \beta_m\boldsymbol{y}^q). \tag{4}$$

Where $m = 1, \ldots, N$ is the candidate ID. $\alpha_m$ and $\beta_m$ are integers that define different rational rotation trials for candidate $m$, and only one of them can be zero. We sort $N$ candidate vector pairs by $\max(|\boldsymbol{u}_m^p|, |\boldsymbol{u}_m^q|)$ such that the first and second minimum length lattice orientation pairs $\{\boldsymbol{u}_{1\mathrm{st}}^p, \boldsymbol{u}_{1\mathrm{st}}^q\}$ and $\{\boldsymbol{u}_{2\mathrm{nd}}^p, \boldsymbol{u}_{2\mathrm{nd}}^q\}$ could be obtained, which are assigned to $\{\boldsymbol{w}^p, \boldsymbol{w}^q\}$ and $\{\boldsymbol{v}^p, \boldsymbol{v}^q\}$, respectively.

In so doing, the irreducible representation $\{\boldsymbol{P}_{\mathrm{IR}}, \boldsymbol{Q}_{\mathrm{IR}}\}$ of a CSL GB character should be written as its transformation matrices plus the first and second minimum length lattice orientations that regulate the irreducible GB supercell:

$$\boldsymbol{P}_{\mathrm{IR}} = \begin{bmatrix} \boldsymbol{x}^p \\ \boldsymbol{y}^p \\ \boldsymbol{z}^p \\ \boldsymbol{w}^p \\ \boldsymbol{v}^p \end{bmatrix} = \begin{bmatrix} H_x^p & K_x^p & L_x^p \\ H_y^p & K_y^p & L_y^p \\ H_z^p & K_z^p & L_z^p \\ H_w^p & K_w^p & L_w^p \\ H_v^p & K_v^p & L_v^p \end{bmatrix}, \tag{5.1}$$

$$\boldsymbol{Q}_{\mathrm{IR}} = \begin{bmatrix} \boldsymbol{x}^q \\ \boldsymbol{y}^q \\ \boldsymbol{z}^q \\ \boldsymbol{w}^q \\ \boldsymbol{v}^q \end{bmatrix} = \begin{bmatrix} H_x^q & K_x^q & L_x^q \\ H_y^q & K_y^q & L_y^q \\ H_z^q & K_z^q & L_z^q \\ H_w^q & K_w^q & L_w^q \\ H_v^q & K_v^q & L_v^q \end{bmatrix}. \tag{5.2}$$

The 2D irreducible size $S_{\mathrm{IR}}$ of the GB supercell is defined by:

$$S_{\mathrm{IR}} = L_w \times L_v \times \cos(\langle \boldsymbol{w}^p, \boldsymbol{v}^p \rangle) = a^2 \times \max(|\boldsymbol{w}^p|, |\boldsymbol{w}^q|) \times \max(|\boldsymbol{v}^p|, |\boldsymbol{v}^q|) \times \cos(\langle \boldsymbol{w}^p, \boldsymbol{v}^p \rangle). \tag{6}$$

Note that equation (4) and the related operations have ensured condition $\cos(\langle \boldsymbol{w}^p, \boldsymbol{v}^p \rangle) = \cos(\langle \boldsymbol{w}^q, \boldsymbol{v}^q \rangle)$. For a given lattice type, one can use material-independent parameter $S_{\mathrm{miIR}} = S_{\mathrm{IR}} \times a^{-2}$ to evaluate and contrast the irreducible GB supercell for different materials. For the GBs belonging to a given material, Σ-independent parameter $S_{\Sigma\mathrm{iIR}} = S_{\mathrm{IR}} \times a^{-2} \times \Sigma^{-1}$ is a good evaluation option instead of $\Sigma$, since the GBs of the same $\Sigma$ could possess very different supercell sizes.

After the establishment of the algorithm, further examination of its reliability and efficiency is required. The best way is to compare its computation with the published reference results. We choose to test our algorithm in an aluminum GB dataset contributed by Homer, et al. [32] (referred to as the Homer dataset), where the GB supercell is constructed following the theoretical work of Banadaki and Patala [51] (referred to as the Banadaki-Patala method) and the GB structure-energy relationship is presented based on an EAM potential from Mishin, et al. [52]. For simplicity, the algorithm only computes $S_{\Sigma\mathrm{iIR}}$ for 14 random GBs picked in the dataset, and the corresponding GB structure and energy are computed by the LAMMPS [53] code GB-Gen [24] following the same manner in the Homer dataset.

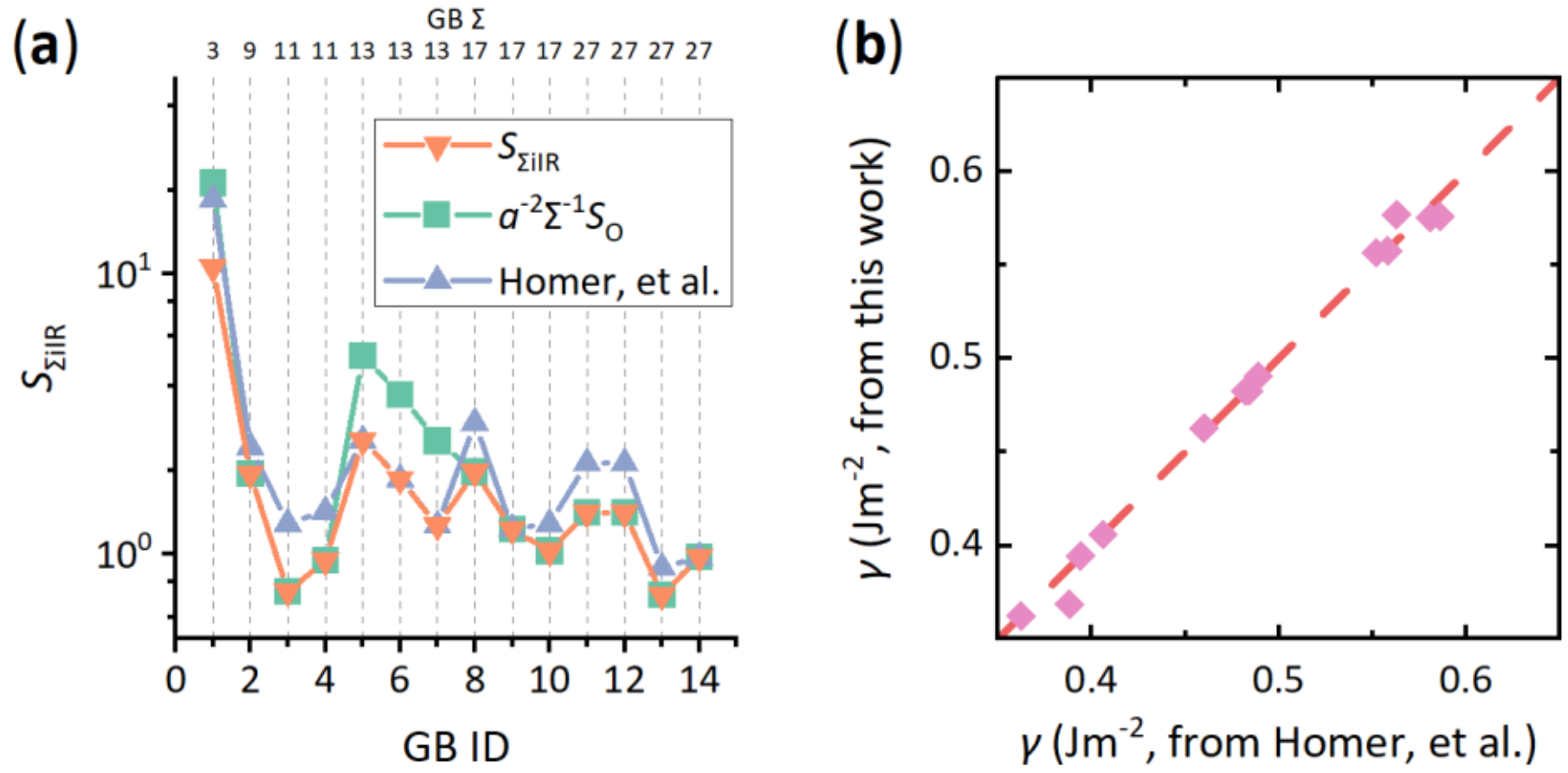

Figure 2. (a) Comparison of computed supercell sizes over a group of 14 random aluminum GBs in the Homer dataset. (b) Parity plot comparing the computed GB energy from this work versus the Homer dataset for the aluminum GBs. The red dashed line denotes perfect data match.

Figure 2a compares the parameter $S_{\Sigma\mathrm{iIR}}$ obtained from the Homer dataset and this work, as well as the counterpart calculated from the orthogonal supercell. Note that different points may coincide if both methods output the same results. It can be seen that this work sometimes computes smaller irreducible supercell sizes than the Banadaki-Patala method employed in the Homer dataset and the orthogonal GB representation. It is likely because this work takes more time to search for potential solutions instead of using the simplest solution of the Diophantine equation in the Banadaki-Patala method. However, this extra time sacrifice in the algorithm calculation is definitely worthwhile, since it is used to provide smaller irreducible cell sizes to trade for saving more computation resources in the atomistic simulation. Figure 2b contrasts the GB energies obtained under two supercell calculation methods. Theoretically, the selection of the supercell only affects the representation, and the GB properties should remain unchanged. The contrast aligns well with the expected result, which indicates that the algorithm is reliable and efficient to determine the irreducible GB supercell.

**Weighted sampling**

Let's assume a scene of high-throughput computation where the user wants to automatically sample and examine the GB character space in whole or in part (examples in Ref. [24, 32, 40]). The examination is limited by a resolution since the user has finite computation resources. Therefore, an automated procedure is required to minimize both computation and human resources spent. We propose a modifiable weighted sampling strategy to achieve this target. As the five dimensions (defined by five macroscopic parameters $p_1, \dots, p_5$) of the GB character space are mutually independent and sampling operation is mathematically similar, we only show the outcome in an arbitrary character dimension $p_c$ where $c$ = 1, …, 5; for example, $p_c$ could be the disorientation angle, the inclination angle [47], and boundary plane normal stereographic projections [50], etc.

Assuming the interval range of $p_c$ is $[0, D_{\text{max}}]$, we construct various $\{\boldsymbol{P}_{\text{T}}, \boldsymbol{Q}_{\text{T}}\}$ and calculate the corresponding parameters $p_c$ and $S_{\Sigma\text{iIR}}$ over that range to form an iteration list. Note that the calculation of $p_c$ from $\{\boldsymbol{P}_{\text{T}}, \boldsymbol{Q}_{\text{T}}\}$ is performed via GB construction/analytical metrics and may differ among different metrics. Then, the iteration list is used to pre-evaluate which GB character therein is relatively "special", based on criterion $S_{\Sigma\text{iIR}} \leq S_{\Sigma\text{iIR}}^{\text{t}}$ (a user-designated threshold), and thus worthy of examination and consideration. Such operation is based on a physical intuition: those GBs with low $S_{\Sigma\text{iIR}}$ values are very likely to possess simple structural units and therefore exhibit special structure-property relationships. An obvious example is that the low Σ GBs are often located in the cusp of the GB energy landscape. We refer to such kind of results as "structural particularity" and will examine it later.

After the evaluation, we assume that the threshold criterion is satisfied by $N_{\text{spec}}$ special GB characters. If the user wants to examine $N_{\text{exam}}$ GB characters and $N_{\text{exam}} \leq N_{\text{spec}}$, then the priority to select should be inversely correlated with $S_{\Sigma\text{iIR}}$ for each character. On the other hand, if $N_{\text{exam}} > N_{\text{spec}}$, all special characters should be examined, which leaves $N_{\text{nons}} = N_{\text{exam}} - N_{\text{spec}}$ quotas for the examination of the rest GBs that are considered not very special. Under such circumstances, we propose a weighted sampling strategy to determine which one should be included in the $N_{\text{nons}}$ characters, to provide an optimal comprehensive coverage of the character dimension space. Given that $N_{\text{exam}}$ characters provide a resolution $Res = D_{\text{max}}/N_{\text{exam}}$, the strategy is to assess how many of the dimension space are already been covered/represented by the special characters, and fill the uncovered space with $N_{\text{nons}}$ characters by weighing the computation cost, particularity and representativeness of the nearby characters. The details of the strategy are stated as the following steps:

(1) Use $Res$ to determine the intervals that are covered and uncovered by the special characters.

(2) Sum the uncovered intervals as $D_{\text{unc}}$, and the local resolution here is $lRes = D_{\text{unc}}/N_{\text{nons}}$.

(3) Calculate the exact parameter $p_c^i = (i - 0.5) \times lRes$ (where $i = 1, \dots, N_{\text{nons}}$) that determines the optimal coverage of the uncovered space for $N_{\text{nons}}$ characters.

(4) Search the $i$-th interval $[p_c^i - lRes/2\,, p_c^i + lRes/2]$ based on the stored $S_{\Sigma\text{iIR}}$ iteration list, to find the $i$-th GB characters with the minimum weighted irreducible supercell size $S_{\text{w}\Sigma\text{iIR}}$.

For GB character $j$ in the $i$-th interval, $S_{\text{w}\Sigma\text{iIR}}$ is computed following:

$$S_{\text{w}\Sigma\text{iIR}}^j = S_{\Sigma\text{iIR}}^j + W\left|p_c^j - p_c^i\right|. \tag{7}$$

Where $W$ is a user-input weight that trades the computational cost with the representativeness.

**Structural particularity**

In this section, the algorithm is applied to a series of aluminum mixed GB characters to predict their structural particularity, and its capability of prediction is examined by atomistic simulations. The mixed GB characters investigated here are geometrically defined following a published construction metric [54], by changing the inclination angle $\varphi$ of the boundary plane between a pair of reference GBs. The metric is briefly described below.

A twist GB with disorientation axis <*H K L*>, disorientation angle $\theta$, and reciprocal density Σ is selected as the first reference, while another tilt GB with the same parameters is selected as the second reference. Such reference selection ensures the boundary plane of the twist GB is perpendicular to the counterpart of the tilt GB. The domain of $\varphi$ is therefore defined as $[0,90°]$, where $\varphi = 0$ gives the first reference twist GB, $\varphi = 90°$ gives the second reference tilt GB, and the mixed GB characters are located at the angular interval $(0,90°)$ via the interpolation equation of the boundary plane normal in the Appendix.

Two pairs of reference GBs are considered, which are (1) $\theta = 36.87°$ Σ5 <001> twist and tilt GB and (2) $\theta = 38.94°$ Σ9 <011> twist and tilt GB. Prior to the GB selection and construction, our algorithm shows the variations of $S_{\Sigma\mathrm{iIR}}$ with $\varphi$ for the Σ5 and Σ9 GBs in Figure 3a1 and 3a2, respectively. It can be seen that the distributions of special GBs with low $S_{\Sigma\mathrm{iIR}}$ values are unique and depend on the selection of disorientation axis and other relevant GB parameters. Based on these pre-evaluation results, we choose to interpolate and investigate $N_{\mathrm{exam}} = 23$ mixed GB characters between each reference pair. So totally 50 GB characters are examined, and their parameters are determined through equation (7) given $p_c = \varphi$, $W = 0.5$ and $D_{\max} = 90°$. Similar to the previous comparisons, the minimum energy structures of these GB characters are obtained at zero temperature and pressure through the LAMMPS code GB-Gen.

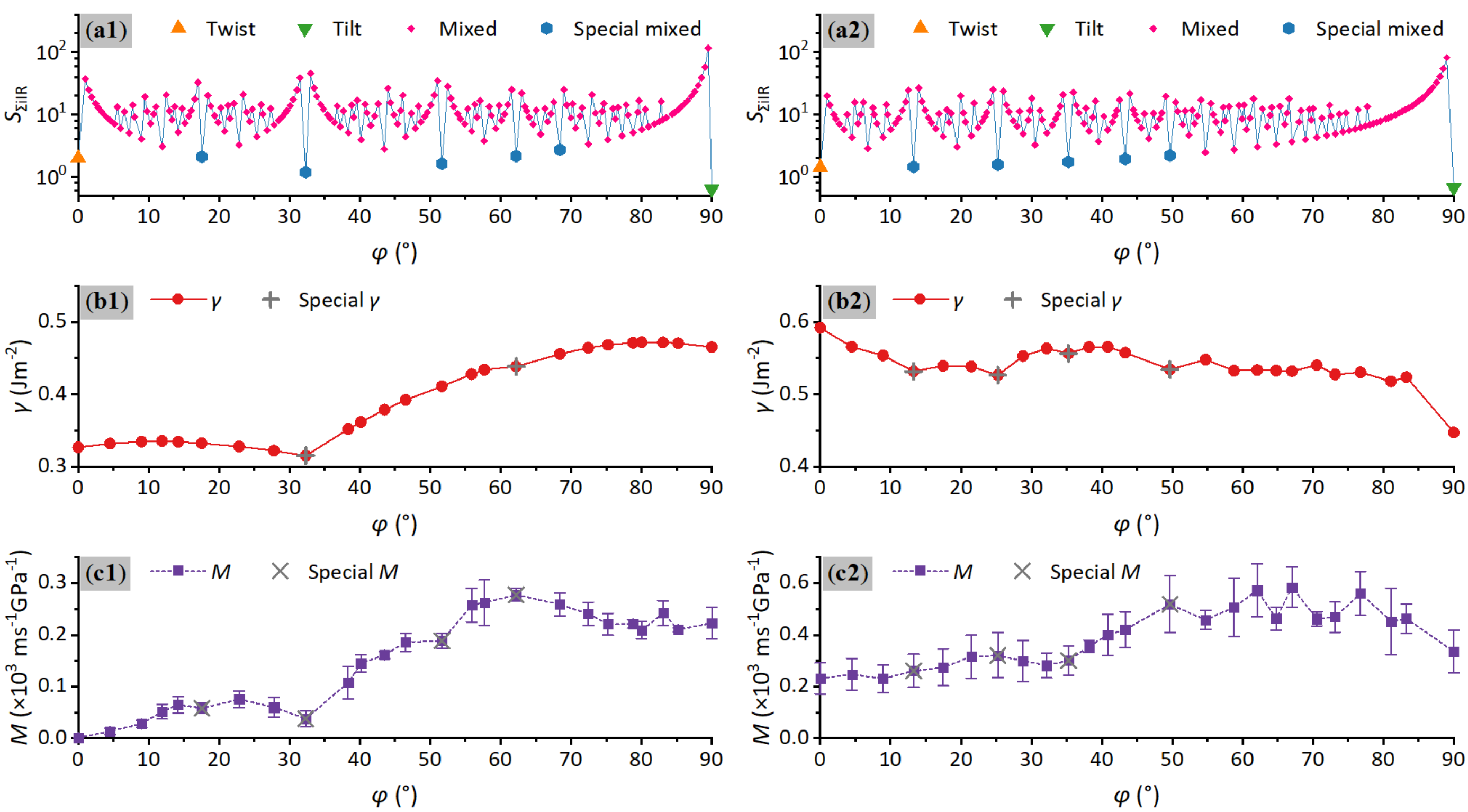

Figure 3. Aluminum GB properties as the function of $\varphi$: (a1) $S_{\Sigma iIR}$ of $\theta = 36.87°$ Σ5 <001> GBs; (b1) $\gamma$ of $\theta = 36.87°$ Σ5 <001> GBs; (c1) $M$ of $\theta = 36.87°$ Σ5 <001> GBs. (a2) $S_{\Sigma iIR}$ of $\theta = 38.94°$ Σ9 <011> GBs; (b2) $\gamma$ of $\theta = 38.94°$ Σ9 <011> GBs; (c2) $M$ of $\theta = 38.94°$ Σ9 <011> GBs.

We first consider the GB structure-energy relationship on the basis of the obtained minimum energy GB structures. Figure 3b1 and 3b2 give the variations of GB energy $\gamma$ with $\varphi$, where the special mixed GBs are found to possess lower $\gamma$ than the nearby mixed GBs and are always located at the local minima of the $\gamma$-$\varphi$ curve. Hence, it is confirmed that the special GBs indicated by the algorithm have certain particularity in terms of the energy. For similar studies about the GB structure-energy relationship, our results reveal that high priority or attention should be given to those special GBs, as they define the overall energy trends.

Subsequently, the finite-temperature GB mobility $M$ (diagonal component of the mobility tensor [55]) is studied. The minimum energy GB structures are relaxed at 300 K for 50 ps using the NPT ensemble [56] to reach thermodynamic equilibrium. Then, the energy-conserving orientational synthetic driving force [57] is applied to induce the GB motion. To avoid the size effects [58] associated with the periodic boundary conditions, the GB structure is replicated in the $x$ and $y$ directions to satisfy the size conditions $L_x > 12a$ and $L_y > 12a$. The GB velocity $s$ normal to the boundary plane is measured at five driving forces ($F$ = 0.025, 0.05, …, 0.125 eV/atom), and $M$ is obtained by fitting the linear equation $s = MF$.

Figure 3c1 and 3c2 plot the mobility trend and use the standard fitting error as the error bar. On the special mixed GBs marked with a grey "×", the mobility trend seems to transit from one variation mode to another. $M$ in these marked data points plays the role of the local minima/maxima compared with the nearby points. However, given that GB mobility is affected more by the atomic interactions than by the GB structures, it is less reflective than the GB energy in terms of the structural particularity. Nevertheless, the influence of structural particularity on the GB mobility still remains recognizable, which merits further exploration in future works.

## Conclusion

An algorithm to calculate the irreducible macroscopic representation for the CSL GB in the cubic lattice is presented in detail. Such representation is applied to establish monoclinic supercells for GB modelling and simulation, allowing access to the complex GB characters. We show that the algorithm calculation is better than the orthogonal representation and a published method. Most importantly, the irreducible supercell size can be used to evaluate the structural particularity of the GB character before constructing its structure, since a special GB character with a lower supercell size is more likely to possess a simpler structural unit. We propose a weighted sampling strategy for the high-throughput simulations to automatically capture such particularity and balance comprehensive coverage of the character space. Finally, with the aid of atomistic simulation, it is shown that the structural particularity is reflected in the structure-property relationships of aluminum mixed GBs in terms of energy and mobility. The special mixed GB characters are often located at the energy cusp and represent the transition between different mobility trends. According to these findings, this algorithm will play its role by telling us which GB is special and should receive our attention.

## Data availability

The presented algorithm has been programmed into the author's private Fortran code "GB-SG.f90", and a compiled executable program of which will be available with other numerical data upon reasonable request.

## Acknowledgements

The author acknowledges the discussions with Prof. E.R. Homer and Prof. J. Han.

## Competing interests

The author declares no competing interests.

## Fundings

This work received NO financial support.

## Author contributions

There is only one author.

## Appendix

The construction metric to interpolate a mixed GB character from a pair of twist and tilt GBs is described here. Given that the GB is located on the *xy* plane, the transformation matrices of a reference twist GB are

$$\boldsymbol{P}_{\mathrm{T}}^{\mathrm{twist}} = \begin{bmatrix} \boldsymbol{x}^p \\ \boldsymbol{y}^p \\ \boldsymbol{z} \end{bmatrix}, \boldsymbol{Q}_{\mathrm{T}}^{\mathrm{twist}} = \begin{bmatrix} \boldsymbol{x}^q \\ \boldsymbol{y}^q \\ \boldsymbol{z} \end{bmatrix}, \quad \text{(A1.1)}$$

and the counterparts of another reference tilt GB are

$$\boldsymbol{P}_{\mathrm{T}}^{\mathrm{tilt}} = \begin{bmatrix} \boldsymbol{x}^p \\ -\boldsymbol{z} \\ \boldsymbol{y}^p \end{bmatrix}, \boldsymbol{Q}_{\mathrm{T}}^{\mathrm{tilt}} = \begin{bmatrix} \boldsymbol{x}^q \\ -\boldsymbol{z} \\ \boldsymbol{y}^q \end{bmatrix}. \quad \text{(A1.2)}$$

Where $\boldsymbol{x}^p\boldsymbol{z} = \boldsymbol{x}^q\boldsymbol{z} = \boldsymbol{y}^p\boldsymbol{z} = \boldsymbol{y}^q\boldsymbol{z} = \boldsymbol{x}^p\boldsymbol{y}^p = \boldsymbol{x}^q\boldsymbol{y}^q = 0$, $|\boldsymbol{x}^p| = |\boldsymbol{x}^q|$, and $|\boldsymbol{y}^p| = |\boldsymbol{y}^q|$. A mixed GB character between the two references is defined as

$$\boldsymbol{P}_{\mathrm{T}}^{\mathrm{mixed}} = \begin{bmatrix} \boldsymbol{x}^{pm} \\ \boldsymbol{y}^{pm} \\ \boldsymbol{z}^{pm} \end{bmatrix} = \begin{bmatrix} \boldsymbol{x}^p \\ \mathrm{red}(f_{\mathrm{s}}|\boldsymbol{z}|^2\boldsymbol{y}^p - f_{\mathrm{r}}|\boldsymbol{y}^p|^2\boldsymbol{z}) \\ \mathrm{red}(f_{\mathrm{s}}\boldsymbol{z} + f_{\mathrm{r}}\boldsymbol{y}^p) \end{bmatrix}, \boldsymbol{Q}_{\mathrm{T}}^{\mathrm{mixed}} = \begin{bmatrix} \boldsymbol{x}^{qm} \\ \boldsymbol{y}^{qm} \\ \boldsymbol{z}^{qm} \end{bmatrix} = \begin{bmatrix} \boldsymbol{x}^q \\ \mathrm{red}(f_{\mathrm{s}}|\boldsymbol{z}|^2\boldsymbol{y}^q - f_{\mathrm{r}}|\boldsymbol{y}^q|^2\boldsymbol{z}) \\ \mathrm{red}(f_{\mathrm{s}}\boldsymbol{z} + f_{\mathrm{r}}\boldsymbol{y}^q) \end{bmatrix}. \quad \text{(A2)}$$

Where $f_{\mathrm{s}}$ and $f_{\mathrm{r}}$ are the rotation scaling factor and rotation factor, respectively. Both are integers designated by the user. In so doing, the boundary disorientation axis is $\boldsymbol{z}$, and parameters $\theta$ and $\varphi$ are calculated by

$$\theta = \arccos\left(\frac{\boldsymbol{x}^p\boldsymbol{x}^q}{|\boldsymbol{x}^p||\boldsymbol{x}^q|}\right), \quad \text{(A3)}$$

and

$$\varphi = \arccos\left(\frac{\boldsymbol{z}\boldsymbol{z}^{pm}}{|\boldsymbol{z}||\boldsymbol{z}^{pm}|}\right) = \arccos\left(\frac{\boldsymbol{z}\boldsymbol{z}^{qm}}{|\boldsymbol{z}||\boldsymbol{z}^{qm}|}\right). \quad \text{(A4)}$$

$\Sigma$ is the odd factor of the term

$$\sqrt{|\boldsymbol{x}^p|^2|\boldsymbol{y}^p|^2|\boldsymbol{z}|^2} = \sqrt{|\boldsymbol{x}^q|^2|\boldsymbol{y}^q|^2|\boldsymbol{z}|^2}. \quad \text{(A5)}$$

Note that for all mixed GB characters interpolated from the same reference pairs, $\theta$ and $\Sigma$ are the same but $\varphi$ is different. To construct $\Sigma5$ and $\Sigma9$ GBs, we employ $\boldsymbol{x}^p = \langle 310 \rangle$, $\boldsymbol{y}^p = \langle \bar{1}30 \rangle$, $\boldsymbol{x}^q = \langle 3\bar{1}0 \rangle$, $\boldsymbol{y}^q = \langle 130 \rangle$, $\boldsymbol{z} = \langle 001 \rangle$, and $\boldsymbol{x}^p = \langle 41\bar{1} \rangle$, $\boldsymbol{y}^p = \langle \bar{1}2\bar{2} \rangle$, $\boldsymbol{x}^q = \langle 4\bar{1}1 \rangle$, $\boldsymbol{y}^q = \langle 12\bar{2} \rangle$, $\boldsymbol{z} = \langle 011 \rangle$, respectively.

Table A1. The aluminum GB characters studied in this work.

| ID | Σ | $\varphi$ | $f_r$ & $f_s$ | $S_{\Sigma iIR}$ | ID | Σ | $\varphi$ | $f_r$ & $f_s$ | $S_{\Sigma iIR}$ |
|---|---|---|---|---|---|---|---|---|---|
| 1 | 5 | 0.00 | 0 & 1 | 0.71 | 1 | 9 | 0.00 | 0 & 1 | 1.41 |
| 2 | 5 | 4.52 | 1 & 40 | 8.02 | 2 | 9 | 4.49 | 1 & 27 | 4.14 |
| 3 | 5 | 8.98 | 1 & 20 | 4.05 | 3 | 9 | 8.93 | 2 & 27 | 4.29 |
| 4 | 5 | 11.90 | 1 & 15 | 2.91 | 4 | 9 | 13.26 | 3 & 27 | 1.20 |
| 5 | 5 | 14.20 | 2 & 25 | 5.06 | 5 | 9 | 17.44 | 4 & 27 | 4.45 |
| 6 | 5 | 17.55 | 1 & 10 | 2.10 | 6 | 9 | 21.44 | 5 & 27 | 4.45 |
| 7 | 5 | 22.86 | 2 & 15 | 3.11 | 7 | 9 | 25.24 | 6 & 27 | 1.56 |
| 8 | 5 | 27.79 | 5 & 30 | 6.78 | 8 | 9 | 28.81 | 7 & 27 | 4.74 |
| 9 | 5 | 32.31 | 1 & 5 | 0.90 | 9 | 9 | 32.15 | 8 & 27 | 5.01 |
| 10 | 5 | 38.33 | 5 & 20 | 5.10 | 10 | 9 | 35.26 | 9 & 27 | 1.50 |
| 11 | 5 | 40.14 | 4 & 15 | 3.80 | 11 | 9 | 38.16 | 10 & 27 | 5.40 |
| 12 | 5 | 43.49 | 3 & 10 | 2.76 | 12 | 9 | 40.83 | 11 & 27 | 5.52 |
| 13 | 5 | 46.51 | 5 & 15 | 4.25 | 13 | 9 | 43.31 | 12 & 27 | 1.94 |
| 14 | 5 | 51.67 | 2 & 5 | 1.37 | 14 | 9 | 49.68 | 15 & 27 | 1.99 |
| 15 | 5 | 55.88 | 7 & 15 | 5.26 | 15 | 9 | 54.74 | 18 & 27 | 2.45 |
| 16 | 5 | 57.69 | 5 & 10 | 3.74 | 16 | 9 | 58.78 | 21 & 27 | 2.56 |
| 17 | 5 | 62.21 | 3 & 5 | 1.94 | 17 | 9 | 62.06 | 24 & 27 | 3.02 |
| 18 | 5 | 68.43 | 4 & 5 | 2.55 | 18 | 9 | 64.76 | 27 & 27 | 3.18 |
| 19 | 5 | 72.45 | 5 & 5 | 3.18 | 19 | 9 | 67.01 | 30 & 27 | 3.62 |
| 20 | 5 | 75.24 | 6 & 5 | 3.80 | 20 | 9 | 70.53 | 36 & 27 | 4.24 |
| 21 | 5 | 78.82 | 8 & 5 | 5.06 | 21 | 9 | 73.14 | 42 & 27 | 4.88 |
| 22 | 5 | 80.04 | 9 & 5 | 5.69 | 22 | 9 | 76.74 | 54 & 27 | 6.16 |
| 23 | 5 | 83.07 | 13 & 5 | 8.22 | 23 | 9 | 81.07 | 81 & 27 | 9.06 |
| 24 | 5 | 85.24 | 19 & 5 | 12.02 | 24 | 9 | 83.28 | 108 & 27 | 12.08 |
| 25 | 5 | 90.00 | 1 & 0 | 0.63 | 25 | 9 | 90.00 | 1 & 0 | 0.67 |